\documentclass[12pt]{article}

\usepackage[paper=letterpaper,margin=0.9in]{geometry}

\begin{document}

\def\CA{{\cal A}}
\def\CB{{\cal B}}
\def\CC{{\cal C}}
\def\CD{{\cal D}}
\def\CE{{\cal E}}
\def\CF{{\cal F}}
\def\CG{{\cal G}}
\def\CH{{\cal H}}
\def\CI{{\cal I}}
\def\CJ{{\cal J}}
\def\CK{{\cal K}}
\def\CL{{\cal L}}
\def\CM{{\cal M}}
\def\CN{{\cal N}}
\def\CO{{\cal O}}
\def\CP{{\cal P}}
\def\CQ{{\cal Q}}
\def\CR{{\cal R}}
\def\CS{{\cal S}}
\def\CT{{\cal T}}
\def\CU{{\cal U}}
\def\CV{{\cal V}}
\def\CW{{\cal W}}
\def\CX{{\cal X}}
\def\CY{{\cal Y}}
\def\CZ{{\cal Z}}

\newcommand{\todo}[1]{{\em \small {#1}}\marginpar{$\Longleftarrow$}}
\newcommand{\labell}[1]{\label{#1}\qquad_{#1}} 
\newcommand{\bbibitem}[1]{\bibitem{#1}\marginpar{#1}}
\newcommand{\llabel}[1]{\label{#1}\marginpar{#1}}

\newcommand{\sphere}[0]{{\rm S}^3}
\newcommand{\su}[0]{{\rm SU(2)}}
\newcommand{\so}[0]{{\rm SO(4)}}
\newcommand{\bK}[0]{{\bf K}}
\newcommand{\bL}[0]{{\bf L}}
\newcommand{\bR}[0]{{\bf R}}
\newcommand{\tK}[0]{\tilde{K}}
\newcommand{\tL}[0]{\bar{L}}
\newcommand{\tR}[0]{\tilde{R}}

\newcommand{\ack}[1]{[{\bf Ack!: {#1}}]}

\newcommand{\btzm}[0]{BTZ$_{\rm M}$}
\newcommand{\ads}[1]{{\rm AdS}_{#1}}
\newcommand{\ds}[1]{{\rm dS}_{#1}}
\newcommand{\dS}[1]{{\rm dS}_{#1}}
\newcommand{\eds}[1]{{\rm EdS}_{#1}}
\newcommand{\sph}[1]{{\rm S}^{#1}}
\newcommand{\gn}[0]{G_N}
\newcommand{\SL}[0]{{\rm SL}(2,R)}
\newcommand{\cosm}[0]{R}
\newcommand{\hdim}[0]{\bar{h}}
\newcommand{\bw}[0]{\bar{w}}
\newcommand{\bz}[0]{\bar{z}}
\newcommand{\be}{\begin{equation}}
\newcommand{\ee}{\end{equation}}
\newcommand{\bea}{\begin{eqnarray}}
\newcommand{\eea}{\end{eqnarray}}
\newcommand{\pat}{\partial}
\newcommand{\lp}{\lambda_+}
\newcommand{\bx}{ {\bf x}}
\newcommand{\bk}{{\bf k}}
\newcommand{\bb}{{\bf b}}
\newcommand{\BB}{{\bf B}}
\newcommand{\tp}{\tilde{\phi}}
\hyphenation{Min-kow-ski}

\newcommand{\pa}{\partial}
\newcommand{\eref}[1]{(\ref{#1})}

\def\apr{\alpha'}
\def\str{{str}}
\def\lstr{\ell_\str}
\def\gstr{g_\str}
\def\Mstr{M_\str}
\def\lpl{\ell_{pl}}
\def\Mpl{M_{pl}}
\def\varep{\varepsilon}
\def\del{\nabla}
\def\grad{\nabla}
\def\tr{\hbox{tr}}
\def\perp{\bot}
\def\half{\frac{1}{2}}
\def\p{\partial}
\def\perp{\bot}
\def\eps{\epsilon}

\newcommand{\BC}{\mathbb{C}}
\newcommand{\BR}{\mathbb{R}}
\newcommand{\BZ}{\mathbb{Z}}
\newcommand{\bra}[1]{\langle{#1}|}
\newcommand{\ket}[1]{|{#1}\rangle}
\newcommand{\vev}[1]{\langle{#1}\rangle}
\newcommand{\Real}{\mathfrak{Re}}
\newcommand{\Imag}{\mathfrak{Im}}
\newcommand{\talpha}{{\widetilde{\alpha}}}
\newcommand{\Ham}{{\widehat{H}}}
\newcommand{\al}{\alpha}
\newcommand\x{{\bf x}}
\newcommand\y{{\bf y}}

\def\NPB{{\it Nucl. Phys. }{\bf B}}
\def\PL{{\it Phys. Lett. }}
\def\PRL{{\it Phys. Rev. Lett. }}
\def\PRD{{\it Phys. Rev. }{\bf D}}
\def\CQG{{\it Class. Quantum Grav. }}
\def\JMP{{\it J. Math. Phys. }}
\def\SJNP{{\it Sov. J. Nucl. Phys. }}
\def\SPJ{{\it Sov. Phys. J. }}
\def\JETPL{{\it JETP Lett. }}
\def\TMP{{\it Theor. Math. Phys. }}
\def\IJMPA{{\it Int. J. Mod. Phys. }{\bf A}}
\def\MPL{{\it Mod. Phys. Lett. }}
\def\CMP{{\it Commun. Math. Phys. }}
\def\AP{{\it Ann. Phys. }}
\def\PR{{\it Phys. Rep. }}


\newcommand{\extd}{{\rm d}}

\def\be{ \begin{equation}}
\def\ee{\end{equation}}
\def\bes{\begin{eqnarray}}
\def\ees{\end{eqnarray}}

\renewcommand{\thepage}{\arabic{page}}
\setcounter{page}{1}

\rightline{VPI-IPNAS-07-10}

\vskip 0.75 cm
\renewcommand{\thefootnote}{\fnsymbol{footnote}}
\centerline{\Large \bf On the spectrum of pure Yang-Mills theory\footnote{To appear in
the proceedings of the first Sowers Theoretical Physics workshop, Virginia Tech, May 2007.}}
\vskip 0.75 cm

\centerline{{\bf
Laurent Freidel,${}^{1,2}$\footnote{lfreidel@perimeterinstitute.ca}
Robert G. Leigh${}^{3}$\footnote{rgleigh@uiuc.edu},
Djordje Minic${}^{4}$\footnote{dminic@vt.edu}
and Alexandr Yelnikov${}^{5}$\footnote{yelnikov@yahoo.com}
}}
\vskip .5cm
\centerline{${}^1$\it Perimeter Institute for Theoretical Physics,}
\centerline{\it 31 Caroline St. N. Waterloo, N2L 2Y5, ON, Canada. }
\vskip .5cm
\centerline{${}^2$\it Laboratoire de Physique, Ecole Normale Sup{\'e}rieure de Lyon}
\centerline{\it 46 All{\'e}e d'Italie, 69364 Lyon, Cedex 07, France.}
\vskip .5cm
\centerline{${}^3$\it Department of Physics,
University of Illinois at Urbana-Champaign}
\centerline{\it 1110 West Green Street, Urbana, IL 61801-3080, USA}
\vskip .5cm
\centerline{${}^4$\it Institute for Particle, Nuclear and Astronomical Sciences, Department 
of Physics, }
\centerline{\it Virginia Tech, Blacksburg, VA 24061, U.S.A.}
\vskip .5cm
\centerline{${}^5$\it Physics Department, City College of the City University of New York}
\centerline{\it 160 Convent Avenue, New York, NY 10031, U.S.A.}

\setcounter{footnote}{0}
\renewcommand{\thefootnote}{\arabic{footnote}}

\begin{abstract}
In this note we discuss the wave functional approach to the
spectrum of pure Yang-Mills theory in $2+1$ and $3+1$ dimensions by
highlighting the issues of dynamical mass generation
and the role played by the kinetic term. 
We extrapolate our recent analysis of 
$2+1$ Yang-Mills theory to $3+1$ dimensions, 
and under certain heuristic assumptions
use a simple quasi-Gaussian 
vacuum wavefunctional for $3+1$ Yang-Mills theory for a preliminary
study of
the spectrum of glueballs which nicely fits the available lattice simulations.

\end{abstract}

\newpage

\section{Introduction and summary}

In this note we address the question of how to
understand the spectrum of pure Yang-Mills (YM) theory using the formalism of
wave-functionals.
In particular we review and clarify our previous work on $2+1$ YM theory \cite{prl} and 
discuss some preliminary results regarding 
the $3+1$ dimensional problem.

The wave functional
approach we have recently used in the context of $2+1$ YM theory
\cite{prl} turns out to fit very nicely the 
existing lattice simulations \cite{teper}.
After reviewing this work we discuss a particularly simple quasi-Gaussian ansatz for the 
vacuum wave-functional which
can be used for a study of the spectrum
of $3+1$ Yang-Mills theory. This ansatz is based on some
heuristic postulates motivated by existing lattice simulations.
Note that even though we cannot argue from first principles that we understand the large N 
limit, we do compare our results to the large N lattice data.
It turns out that the comparison with finite N results does not alter the final results too much.
We follow the general strategy towards a wave-functional approach to the
spectrum of $3+1$ YM that was outlined in
\cite{L}.
The lattice results we compare to are collected, for example, 
in \cite{Lucini:2004my, T, Meyer:2004gx, Morningstar:1999rf, Kuti:1998rh, Hu:1996ys}.


The main physics issues to be addressed in $3+1$ YM theory, as opposed to its
$2+1$ dimensional counterpart are:

1) dynamical generation of a scale independent mass: the so called gap 
equation. We discuss this issue by analyzing the origin of the
mass gap from the geometry of the configuration space in the context
of $2+1$ YM. We then present some very heuristic
arguments about the gap origin in the more difficult $3+1$ dimensional problem.

2) the IR asymptotics of the quasi-Gaussian vacuum wave functional 
and the issue of the evaluation of the string tension
of large planar Wilson loops.

3) finally, there is the question of how to recover the main feature of 
the UV physics, i.e. asymptotic freedom,
the hallmark universal feature of QCD
perturbation theory (as exemplified in the famous $11/3$ one-loop $\beta$ function coefficient).
This turns out to follow from the proper UV asymptotics of the vacuum wave-functional.

By contrast, one does not have to worry about these issues in
the analysis of $2+1$ YM. In that case the theory has a natural dimensionful parameter,
the YM coupling.
So, in our attempt to treat some phenomenological
features of the $3+1$ YM spectrum by using
the experience with $2+1$ YM theory, we implicitly assume that 
the above difficult issues can be addressed in a self-consistent
manner in the context of simple quasi-Gaussian vacuum wave functionals.
For example, we simply assume that 
the mass gap is self-consistently generated
in the wave functional approach to the $3+1$ 
YM theory. Also, the form of the quasi-Gaussian ansatz for the vacuum wave functional
in the $3+1$ dimensional context 
is essentially motivated by the $2+1$ dimensional counterpart. 
The IR asymptotics of the $3+1$ YM theory is very difficult to analyze
from first principles:
see, for example, the discussion in \cite{kogan}.
In the $3+1$ problem, the asymptotic freedom issue is dealt with
by appealing to the correct UV limit of
the vacuum wave functional, the requirement of gauge symmetry
(so that all interaction vertices are correctly captured) and the standard RG
analysis of wave functionals \cite{djnair}.

Our appeal for a self-consistent generation of the mass gap and the
IR asymptotics on the level of the vacuum wave functional does have
some resonance with the philosophy pursued in \cite{migdal} as well
as the AdS/QCD approach \cite{brodsky}. 
The details appear to be very different though.
The essential 
physics ingredient in our case is the constituent picture
which seems to work so nicely in $2+1$ YM \cite{prl}
(for previous attempts at the constituent picture in the 
$3+1$ YM consult, for example, \cite{constituent})

This note is organized as follows:

1) We start with a general discussion of how to understand the spectrum of pure Yang Mills theory 
in the formalism of vacuum wave functionals. We concentrate on here 
$2+1$ dimensional YM theory.
The crucial heuristic points in this approach are the issues of a dynamical 
mass gap generation and the role of the kinetic term.

2) We then present a more phenomenologically minded 
discussion of the spectrum of $3+1$ dimensional
Yang Mills theory based on a quasi-Gaussian vacuum
wave functional. 
This phenomenologically minded ansatz is
on one hand motivated by the results in $2+1$ dimensions and
on the other hand by good numerical comparison with the lattice data.
We think that the results presented in this paper justify the simple quasi-Gaussian vacuum 
wave-functional approach in spite of a somewhat tentative nature of our theoretical analysis.

\section{Wave functionals and the kinetic term}

Here we outline a general procedure regarding the derivation of a vacuum wave functional
for pure YM theory,
which in the case of $2+1$ YM theory happens to be in excellent agreement with the corresponding lattice
data concerning the spectrum of the vacuum states of the theory \cite{prl}.

We contrast the abelian case, which can
be exactly solved, and the more difficult non-abelian case. We do not utilize any special 
variables. The approach is, in principle, very general and transparent. For concreteness, in 
what follows, we concentrate on the $2+1$ dimensional physics.
In later sections we will discuss the $3+1$ dimensional YM theory.

Given the gauge theory Hamiltonian
\be
H = \frac{1}{2}\int (g^2 E_i^2 + \frac{1}{g^2} B^2)
\ee
where $E_i$ and $B$ are the electric and magnetic components of
$F_{\mu \nu} \equiv \partial_{\mu} A_{\nu} - \partial_{\nu} A_{\mu} + [A_{\mu}, A_{\nu}]$,
we wish to reason out the vacuum wave-functional using 
the ordinary connection variables $A^a$. In the Hamiltonian gauge $A_0=0$
\be
H = \frac{1}{2}\int (-g^2 (\frac{\delta}{\delta A_i})^2 + \frac{1}{g^2} B^2)
\ee
We are looking for $\Psi_0$ such that $H \Psi_0 =0$, after subtracting the infinite
vacuum energy term.

In the abelian case the answer is obviously Gaussian. In the non-abelian case
this is not so, for reasons of non-abelian gauge symmetry. So, in $2+1$ dimensions we consider 
a quasi-Gaussian ansatz, which is compatible with the Gauss law
\be
\Psi_0=\exp(- c\int B\left( K(L)\right) B)
\ee
where $c$ is the appropriate constant $1/(2g^2 m)$, $B$ is 
the magnetic field and $L\equiv\frac{\Delta}{4m^2}$,
where $\Delta$ is the covariant Laplacian, and $m$ is 
the appropriate mass parameter. Note that the parameter
$m$ can be explicitly evaluated using the approach of ``corner variables'' \cite{bars}
of Karabali and Nair \cite{knair}. In that case there is a very precise relation (for large $N$)
\be
m \equiv \frac{g^2 N}{2 \pi}
\ee
It suffices to say here that the parameter $m$ is ultimately related to the value of the
string tension for large Wilson loops.
Also, at least formally
\be
K(L) = \sum_{n=0}^{\infty} c_n L^n
\ee
and thus more generally
\begin{equation}\label{eq:SchrEqGen}
{\cal H}\Psi_0=E_0\Psi_0=\left[ E_0+\int B^a \left( {\cal R}\right)B^a +\ldots\right]\Psi_0 .
\end{equation}
The (divergent) vacuum energy $E_0$ can be isolated and as expected, the leading divergence in the UV is 
cubic. Next, what we need to do is compute the expression that we have labeled by ${\cal R}$ and set 
it to zero. This will constitute an equation for the kernel $K$.

Let us consider the various terms in ${\cal R}$. First, the potential term $B^a B^a$ clearly contributes 
a fixed constant to ${\cal R}$.
The remainder of ${\cal R}$ will come from the action of the kinetic energy operator $T_{KN}$. Given 
the form of the vacuum wave functional $\Psi_0 \sim e^P$, it is elementary
to derive
\be
\frac{\delta\Psi_0}{\delta A^a(z)}=\frac{\delta P}{\delta A^a(z)}\Psi_0
\ee
and thus
\be\label{eq:formalSch}
\frac{\delta^2\Psi_0}{\delta A^a(z)A^b(w)}=\left[\frac{\delta^2 P}{\delta A^a(z)A^b(w)}+\frac{\delta P}{\delta A^a(z)}\frac{\delta P}{\delta A^b(w)}\right]\Psi_0
\ee
from which we find
\begin{equation}\label{eq:PSchrodinger}
{\cal H}\Psi_0 = \left[T P+ m\int_{z,w} \frac{\delta P}{\delta A^a(z)}\frac{\delta P}{\delta A^b(w)} + \frac{1}{ m} 
\int B^a B^a\right]\Psi_0.
\end{equation}
The second term in brackets is easy to compute. Since we want to solve Schr\"odinger equation 
to quadratic in $B^a$ order only
\begin{equation}\label{eq:Pquadratic}
P \sim
\int B^a \ K\left(\frac{\Delta}{m^2}\right) B^a + \ldots
\end{equation} 
By evaluating functional derivatives $\delta P/\delta A$ we see  that the second term in brackets in (\ref{eq:PSchrodinger}) is equal to
\begin{equation}
\int B^a \left[ \frac{\Delta}{m^2} K^2\left(\frac{\Delta}{m^2}\right)\right] B^a
\end{equation}
Therefore, we deduce that this term contributes $L K^2(L)$ into $\cal R$.
This result also holds in the pure Abelian case.

Let us finally proceed to the $T P$ term in (\ref{eq:PSchrodinger}). 
In the Abelian case this term does not depend on field variables and gives (infinite) vacuum energy only. In the non-Abelian case
the kinetic term plays a very important role.
In $2+1$ dimensions the simplest way to proceed is to assume that the
kinetic energy operator $T$ acts homogeneously on any 
local operator-valued function of $B$. This is natural in the holomorphic
corner variable approach of Karabali, Kim and Nair. 
This assumption can be somewhat justified by explicit computations
involving the action of $T$ on the potential term $B^2$. 
In that particular case the action of $T$ would simply 
produce $2m P$, since $P$ is quadratic in $B$. 
Now, by extrapolating this statement into \cite{prl}
\begin{equation}\label{eq:TOn}
T\CO_n = (2+n)m\, \CO_n+\ldots
\end{equation}
where $\CO_n = B L^n B$, one gets much more information\footnote{A possible calculational justification for this statement might be obtained by following
the strategy proposed in the recent work by Fukuma et al \cite{fukuma}.}.
Notice that here ellipsis stand for terms of higher order in $B$, but the same mass 
dimension as $\CO_n$, and which mix with $\CO_n$ under the action
of $T$. 
Finally, note that this statement in some sense incorporates the role of topology
(i.e. compactness of the configuration space)
in the non-abelian case, and 
encapsulates the importance of the kinetic term in 
the dynamical origin of the mass gap, which is intuitively very much in line with \cite{F}.

Returning to our discussion of the Schr\"odinger equation we note that (\ref{eq:TOn}) implies
\be
T P=- c \int B \left\{\frac{1}{2}\sum_n c_n(2+n)L^n \right\} B.
\ee
It is convenient to write the factor in braces formally as
\begin{equation}
\frac{1}{2 L} \frac{d}{dL}\left[ L^2K(L)\right].
\end{equation}
Assembling all of these results, we then find the following Riccati equation \cite{prl}
\begin{equation}\label{eq:Riccati}
{\cal R}\sim \left[-K - \frac{L}{2} \frac{d}{dL}[K(L)] + L K^2 +1\right]=0.
\end{equation}

Note that the abelian answer is given by the last algebraic part of the equation, which does
not have anything to do with the spectrum of the kinetic term, i.e. $L K^2 +1 =0$.
This of course, has to be reproduced even in the non-abelian situation at UV.
Now, the discrete spectrum of T is essential for the emergence of the gap.
So the first part of the equation $\frac{1}{2 L} \frac{d}{dL}\left[ L^2K(L)\right]$ is crucial. In the 
IR limit only the factor $K+1=0$ survives.
The derivative term $\frac{L}{2} \frac{d}{dL}[K(L)]$ interpolates between the UV and the IR
limits.
The other essential ingredient is that there is only one normalizable $K$ that solves
the whole differential equation, as discussed in what follows.

Finally, note that the whole discussion is in principle valid for any rank of the gauge group.
Even though we do not have a very good reason to rely on large N, we will perform the actual
comparison of analytic results for the spectrum with large N lattice data (which are
already very close to the $SU(3)$ case.)

\subsection{The Gluonic Kernel}

Although the above Riccati equation is non-linear, it is easily transformed into a linear 
second-order equation of the Bessel type, and one finds a general solution of the form
\be
K(L)=\frac{1}{\sqrt{L}}\frac{CA_2(4\sqrt{L})+Y_2(4\sqrt{L})}{CA_1(4\sqrt{L})+Y_1(4\sqrt{L})}
\ee
where $C$ is a constant and $J_n$ ($Y_n$) denote the Bessel functions of the first (second) kind.
As explained in \cite{prl} it is remarkable that the only normalizable wave functional is 
obtained for $C\to \infty$, which is also the only case that has both the correct UV behavior 
appropriate to asymptotic freedom, as well as the correct IR behavior appropriate to confinement 
and a mass gap!
This solution is of the form
\begin{equation}\label{eq:normvacuum}
K(L) = \frac{1}{\sqrt{L}} \frac{J_2(4 \sqrt{L})}{J_1(4 \sqrt{L})}
\end{equation}
This remarkable formula is reminiscent of similar results in related
contexts \cite{migdal}; here, it encodes information on the spectrum of the theory. We note 
that this kernel has the following asymptotics
(where $L \sim -\vec p^2/4m^2$)
\begin{equation}
p \to 0, \quad K \to 1;\ \ \ \ \   p \to \infty, \quad K \to 2m/p
\end{equation}
consistent with confinement and asymptotic freedom, respectively (Note that the argument of 
Bessel functions is imaginary;
so instead of $J_n$ we have $I_n$ Bessel functions.)

Now using standard Bessel function identities we may expand 
\begin{equation}
\frac{J_{1}(u)}{J_{2}(u)} = \frac{4}{u} + 2 u \sum_{n = 1}^{\infty}\, \frac{1}{u^2 - \gamma_{2,n}^2}
\end{equation}
where the $\gamma_{2,n}$ are the ordered zeros of $J_2(u)$ \cite{prl}. 
The inverse kernel is thus 
\begin{equation}
K^{-1}(L) = \sqrt{L}\frac{J_{1}(4 \sqrt{L})}{J_{2}(4 \sqrt{L})}  
= 1 + 8L\sum_{n = 1}^{\infty}\,
 \frac{1}{16 L - \gamma_{2,n}^2}.
\end{equation}
Now, if we regard $L\simeq \pa\bar\pa /m^2$, in terms of momentum $p$ we find
\begin{equation}\label{eq:KernelInverseK}
K^{-1}(p) =  1 + \frac{1}{2} \sum_{n = 1}^{\infty}\, \frac{\vec p^2}{\vec p^2 + M_{n}^2}.
\end{equation}
Here 
\begin{equation}\label{constmass}
M_n = \frac{\gamma_{2,n} m}{2}.
\end{equation}
$M_n$'s can be interpreted as constituents out of which glueball masses are constructed. 
It is not difficult now to find a Fourier transform of inverse kernel $K^{-1}(k)$. By 
rewriting (\ref{eq:KernelInverseK}) as 
\begin{equation}
K^{-1}(p) =  1 + \frac{1}{2} \sum_{n = 1}^{\infty}\, \left(1-\frac{M_{n}^2}{\vec p^2 + M_{n}^2}\right)
\end{equation}
we immediately obtain
\begin{equation}\label{eq:KFourier}
K^{-1}(x-y) = \delta^{(2)}(x-y) + \frac{1}{2} \sum_{n = 1}^{\infty}\, \left( \delta^{(2)}(x-y) - \frac{M_n^2}{2\pi}\, K_0(M_n |x-y|)  \right)
\end{equation}
where $K_0(x)$ is the modified Bessel function of the third kind. At asymptotically large spatial 
separations $|x-y| \to \infty$ this takes the form
\begin{equation}\label{eq:Kpossp}
K^{-1}(|x-y|) \approx -\frac{1}{4\sqrt{2\pi |x-y|}} \sum_{n = 1}^{\infty} (M_{n})^{\frac{3}{2}}  e^{-M_n |x-y|} .
\end{equation}
As shown in \cite{prl} we can treat $B$ as a free
field as well and therefore
\begin{equation}
\langle  B(x)\  B(y)\rangle \sim  K^{-1}(x-y).
\end{equation}
The constituent masses are controlled by the zeros of $J_2$.
\be
M_n = \frac{ \gamma_{2,n} m}{2}
\ee
The glueball masses then follow by evaluating
the correlators of the relevant operators that 
probe the glueball states. For example, the mass of the $0^{++}$ glueball may be probed
by the operator $Tr B^2$. We have \cite{prl}
\begin{equation}
\langle {{\rm Tr}\,(B^2)}_x \
{{\rm Tr}\,(B^2)}_y\rangle \sim \left( K^{-1}(|x-y|)\right)^2.
\end{equation}
and thus the $0^{++}$ mass is \cite{prl}
\begin{equation}\label{0++mass}
M_{0^{++}} = M_1 + M_1
\end{equation}
A more complete discussion of other glueball masses can be found in \cite{prl}.

Now it is also not difficult to derive expression for the vacuum expectation of the 
large Wilson loop of area 
$A$ . At large $N$ 
\begin{equation}
\langle \Phi(C)\rangle = \langle {\rm Tr}\, P\, {\rm exp}\left( \oint_C A\right)  \rangle \ \ \stackrel{N\to\infty}{\longrightarrow} \ \
N\, {\rm exp}\left( - N \int {d^2x} {d^2y}\ \langle B(x) B(y)\rangle \right)
\end{equation}
or
\begin{equation}
\frac{1}{N} \langle\Phi(C)\rangle = {\rm exp} \left(-\frac{g_{YM}^4 N^2}{8\pi} \int d^2x d^2y\, K^{-1}(x-y)\right)
\end{equation}
and from this expression we see that leading $\delta$-function gives area law
with a string tension that fits beautifully the lattice simulations. As for the rest of terms that appear
in (\ref{eq:KFourier}), we may notice that 
\begin{equation}
\int_A d^2x  \left( \delta^{(2)}(x-y) - \frac{M_n^2}{2\pi}\, K_0(M_n |x-y|)  \right)\ \to\ 0 \quad {\rm as} \quad A\to\infty
\end{equation}
and therefore these terms will give corrections to the area law behavior which vanish for asymptotically 
large loops. 
The string tension then follows to be \cite{knair}
\be\label{eq:stringtension}
\sqrt\sigma = \frac{g^2 N}{\sqrt{8 \pi}} = \sqrt{\frac{\pi}{2}}\ m 
\ee
Given this expression we can easily convert glueball masses (\ref{0++mass}) into the units of the square root of the
string tension which makes comparison to lattice data straightforward.

Finally, a couple of comments.
Note that the gluonic wave functional leads to an effective partition function of a pure 2d YM 
theory in
the IR limit. 
Also the purely Gaussian vacuum wave-functionals suffice for self-consistent calculations of
both the glueball masses and the string tension.
Similarly, note that the $J$ constituents do not appear as asymptotic states. 
Thus, it might be plausible that the glueball constituents are ``seeds'' for constituent quarks
once the fermionic degrees of freedom are included.
If so, quark confinement would be ``seeded'' by the confinement mechanism in the
pure glue sector.

\subsection{Generalizing the kernel equation}

One of the 
most intriguing results of \cite{prl} is the
prediction of the $0^{++}$ glueball mass which agrees extremely well
with the lattice simulations. The derivation
of that result crucially
depends on equation (12) which leads to nontrivial kernel equation (\ref{eq:Riccati}). In the absence
of a rigorous derivation of that equation from first principles,
we want to show that equation (12)
is well motivated phenomenologically. To this end we
may consider a modification of equation (12). We
will also use the following results based on such a modified kernel equation
to motivate our phenomenologically minded study of the
spectrum of $3+1$ YM theory.

Suppose that the action of the kinetic term is as follows\footnote{In principle we could have taken an even more 
general expression $T \CO_n = (a +bn) m \CO_n +...$ . This modification, however, does not lead to any new conclusions
compared to (\ref{generalized}).}
\be\label{generalized}
T \CO_n = (2 +bn) m \CO_n +...
\ee
where $b$ is some free parameter. 

Then the kernel $K(L)$ satisfies the following equation of the Riccati type
\begin{equation}\label{Riccati}
-K - b\frac{L}{2} \frac{d}{dL}[K(L)] +  L K^2 +1 =0
\end{equation}
After changing variables as
\be
K = - b\frac{y'}{2y}
\ee
where the prime denotes the derivative with respect to $L$, we
arrive at
\be
y'' + \frac{2}{bL} y' + \frac{4}{b^2 L}y =0
\ee
which is readily solved in terms of Bessel functions.
The general solution is
\begin{equation}
y = c_1 L^{\frac{1 - 2/b}{2}} J_{2/b -1} (4/b \sqrt{L}) + c_2 L^ 
{\frac{1 - 2/b}{2}} Y_{2/b -1} (4/b \sqrt{L}).
\end{equation}
The physically relevant solution will be
\be
y =  L^{\frac{1 - 2/b}{2}} J_{2/b -1} (4/b \sqrt{L})
\ee
Using the standard recursive formulae
\begin{equation}\label{eq:besselrelns}
\begin{array}{ccl}
J_{\nu - 1}(u) + J_{\nu + 1}(u) &=& \frac{2 \nu}{u}J_{\nu}(u) \\
&&\\
J_{\nu - 1}(u) - J_{\nu + 1}(u) &=& 2 J'_{\nu}(u)
\end{array}
\end{equation}
and analogous formulae for $Y_\nu$, we arrive at the general solution 
\begin{equation}
K(L)= \frac{1}{\sqrt{L}}\frac{C J_{2/b}(4/b \sqrt{L}) + 
Y_{2/b}(4/b \sqrt{L})}{C J_{2/b-1}(4/b \sqrt{L}) + Y_{2/b-1}(4/b \sqrt{L})}.
\end{equation}
Once again, the only solution which is normalizable in UV and IR 
is given by
\begin{equation}
K(L) = \frac{1}{\sqrt{L}}\frac{J_{2/b}(4/b \sqrt{L})}{J_{2/b-1}(4/b \sqrt{L})}.
\end{equation}
We see that the argument (and order) of the relevant Bessel  
functions depends on $b$, and
we recover the standard $2+1$ result if $b=1$.
Note that in general if
$T O_n = (a +bn) O_n$
we have
\begin{equation}
K(L) = \frac{a}{2} \frac{1}{\sqrt{L}}\frac{J_{a/b}(4/b \sqrt{L})}{J_{a/b-1}(4/b \sqrt{L})}.
\end{equation}
For $a=2$, the kernel has the following asymptotics
(where $L \sim -p^2/4m^2$)
\begin{equation}
p \to 0, \quad K \to 1;\ \ \  p \to \infty, \quad K \to 2m/p
\end{equation}
consistent with confinement and asymptotic freedom, respectively.

Now using standard Bessel function identities (see \cite{prl})
or more generally
\begin{equation}
\frac{J_{\nu + 1}(u)}{J_{\nu}(u)} = 2 u \sum_{n = 1}^{\infty}\, \frac 
{1}{u^2 - \gamma_{\nu,n}^2}
\end{equation}
where the $\gamma_{\nu,n}$ are the ordered zeros of $J_{\nu}(u)$.
The inverse kernel is thus in general
\begin{equation}
K^{-1}(L) =  {\sqrt{L}}\frac{J_{2/b-1}(4/b \sqrt{L})}{J_{2/b}(4/b  
\sqrt{L})}.
\end{equation}

Then the constituent masses in this more general case are given by a  
simple formula, again in parallel with \cite{prl}
\begin{equation}\label{constmass}
M_n = \frac{b \gamma_{2/b,n} m}{2} .
\end{equation}
In the case of $2+1$ YM the ratio of the glueball masses that follows from this constituent mass and
the corresponding expression for the square root for the string tension best fits the
lattice data for $b=1$, as discussed in the previous subsection.

\subsection{A recapitulation}

The effect of the kinetic term seems to be of utmost physical importance
(at least in the context of pure $2+1$ YM theory.)
In the abelian case the kinetic term just contributes to the vacuum energy and is not
essential for the determination of the kernel.
In a topologically non-trivial case, i.e. non-abelian theory, the effect of the kinetic term is crucial.
Heuristically, the compactness of the configuration space seems to
imply a discrete ``spectrum'' for the kinetic term, and this drastically changes the nature of the
kernel and
the resulting mass spectrum of the theory.
The mass spectrum is in turn very nicely captured by a simple quasi-Gaussian vacuum wave functional.
Presumably the compact $U(1)$ could be also treated in this manner, but it is not clear
whether a Gaussian ansatz captures the whole physics.
It is possible that instead of $P \sim \int B K(L) B$ one should consider
something non-quadratic yet consistent with the compactness of the configuration space, say
$P \sim \int \cos(c_1 B K(L) B)$, where $c_1$ has the appropriate dimensions for the cosine to make sense.

\section{On the origin of the mass gap in pure YM theory}

Here we present a summary of an intuitive argument of Karabali, Kim and Nair concerning the
mass gap in $2+1$ YM theory \cite{knair}. We then discuss possible generalizations of
this argument for the case of $3+1$ YM theory.

In the Karabali-Nair formalism \cite{knair} one obtains that
the configuration space measure is given by the exponent of the 2d WZW action.
This is the volume element which then also defines a normalizable
inner product of gauge invariant wave functionals
\be
\langle 1|2\rangle_{2+1} = \int d\mu(H) e^{2 c_A S_{WZW}(H)} \Psi_1^*(H)\Psi_2(H)
\ee
This configuration space measure is crucial in the argument concerning the
dynamical origin of the mass gap.
(The emphasis here is on the term {\it dynamical}; that the gap should be proportional to the
dimensionful coupling constant is obvious on dimensional grounds.)

First note that the {\it quadratic} part of $S_{WZW}(H)$ is given by,
in terms of the magnetic field, as
\be
S_{WZW} \sim \frac{1}{4\pi} \int dx^2 B \frac{1}{\nabla^2} B + ...
\ee
where the higher order terms  in $B$ are dropped in the exact expression. 

Now, the intuitive argument of \cite{knair} runs as follows.
Denote by $\Delta E$ and $\Delta B$ the root mean square fluctuations of the
electric and magnetic fields.
By utilizing the canonical commutation relations $[E_i^a, A_j^b] = - i \delta_{ij} \delta^{ab}$,
one gets 
$\Delta E \Delta B \sim k$, where $k$ denotes the momentum.
The estimate for the energy is then
\be
\epsilon = \frac{1}{2} [\frac{g^2 k^2}{\Delta B^2} + \frac{\Delta B^2}{g^2}]
\ee
$g$ being the gauge coupling.
Note that this would give, by minimizing over $\Delta B^2$, $\Delta B^2_{min} \sim g^2 k$ which
then implies $\epsilon \sim k$, i.e. no mass gap.
This is indeed the correct answer for the Abelian theory, but 
incorrect for its non-Abelian cousin.

In the non-Abelian case we need to take care of the non-trivial configuration space
measure (because the physical meaning of the uncertainty relation hinges on implicitly
knowing the inner product), which we have in $2+1$ thanks to \cite{knair}.
The equation for the Gaussian part of this measure implies that 
$\Delta B^2 \sim \pi k^2/C_A$ for small values of the momenta, that is in the IR limit.
Thus, as pointed out in \cite{knair}, even though the energy $\epsilon$ is
minimized around $\Delta B \sim k$, the probability, which is dominated by the measure factor,
is centered around $\Delta B^2 \sim k^2 \pi/c_A$.  Consequently, the energy 
scales as $\epsilon \sim g^2 c_A/2\pi + O(k^2)$ and indicates the presence of a
mass gap in the spectrum, proportional to $g^2 c_A$.
Again, note the crucial role played by the kinetic term in this
intuitive argument.

We wish to emphasize the dynamical nature of this heuristic argument, as
opposed to the final result which could be expected on dimensional grounds.
Obviously the dimensional argument fails in $3+1$, so we discuss some options
regarding the generalization of the
dynamical part of the argument.

For example, one option would be to consider the natural generalization of the
above configuration space measure for $2+1$ YM in the $3+1$ 
dimensional context
\be
{\langle 1|2\rangle}_{3+1} = \int d\mu(\vec{B}) e^{2 c_A S_{3+1}(\vec{B})} 
\Psi_1^*(\vec{B})\Psi_2(\vec{B})
\ee
where to quadratic order in $\vec{B}$, by generalizing the quadratic
result from $2+1$
\begin{equation}
S_{3+1} \sim \frac{m}{g^2} \int dx^3 \vec{B} \frac{1}{\nabla^2} \vec{B} + ...
\end{equation}
The parameter $m$ should represent an RG invariant mass scale and
it should be dynamically determined from a gap equation, perhaps along the lines
of \cite{kogan}.
(This formula is natural form the point of view of the formalism of corner variables 
discussed in \cite{4dpapers}. A different expression for this measure was found in
\cite{ynair}. This might indicate the sensitivity of the $3+1$ dimensional
problem to a choice of variables and the fact that there are no obvious
counterparts to the WZW functional in three dimensions.)
By repeating the above argument based on the uncertainty principle we
get that in this case the gap is proportional to $m$, as expected.

Another option would be to distinguish the $2+1$ and $3+1$ dimensional problems
by appealing to a more dynamical role of the potential term in the
$3+1$ dimensional context.
In this case the intuition about the infrared dynamics of $3+1$ YM theory 
based solely on the experience with $2+1$ YM theory would be
most probably incomplete.

Whatever the precise dynamical mechanism of mass generation is
in $3+1$ YM theory, the crucial question is
to determine the mass gap
dynamically (via a gap equation), and relate it to the universal features of perturbation theory (i.e.
asymptotic freedom and $\Lambda_{QCD}$).
We will not explicitly address these difficult issue here. In the next section, we will simply
postulate that the wave-functional ``knows'' about a stable mass parameter $m$,
which is in turn related to the mass gap as well as $\Lambda_{QCD}$.

\section{A preliminary study of the spectrum of $3+1$ YM}

In this section, we 
want to approach the spectrum of $3+1$ YM theory in the
manner used in the analysis of $2+1$ YM, as summarized in 
the preceding part of the paper. Of course, the two theories are radically different,
as outlined in the introduction. 
Our discussion is necessarily phenomenologically minded, given the difficult
issues concerning the dynamical origin of the mass gap and the IR asymptotics
of the vacuum wave functional.
We will be mainly motivated by the existing lattice simulations in $3+1$ YM and
the success of the simple quasi-Gaussian ansatz for the vacuum wave functional
for $2+1$ YM theory.

As in the case of $2+1$ YM we assume the following ansatz for the vacuum wave functional
\be \label{3+1wf}
\Psi_0=\exp(- \frac{1}{2g^2m}\int B_i\left( K(L)\right) B_i)
\ee
with
a non-trivial kernel $K$ and $L \equiv \frac{\Delta}{4m^2}$, where $\Delta$ is the $3$d
covariant Laplacian. 
Note, that in contrast to $2+1$ YM here it is crucial to determine
the dynamical scale $m$ (via a gap equation), which is not explicitly present in the bare
Hamiltonian of $3+1$ YM.
We will model the kernel $K(L)$ motivated by the discussion from section 2.2 of a generalized kernel
equation for the $2+1$ dimensional problem.
As pointed out in section 2.2, the UV behavior of such a kernel will be just 
what we need in $3+1$ YM theory
\be\label{Psi_UV}
\Psi_0^{\rm UV} = \exp(- \frac{1}{2g^2}\int B_i\left(\frac{1}{\sqrt{\Delta}} \right) B_i + \ldots)
\ee
In that limit the mass $m$ cancels and we are left with the correct UV result
known from QCD perturbation theory.
By applying the standard background field methods to such a UV wave functional
(following, for example, the nice work of
Zarembo in \cite{djnair})
we can readily recover the universal features of perturbation theory, such as the famous
$11/3$ coefficient in the one-loop beta function.
The corrections to this UV result (indicated by "$\ldots$" in (\ref{Psi_UV})) can be organized in powers of the mass $m$ over
the powers of momenta and as such they encapsulate non-perturbative effects.
Given the usual decoupling between the perturbative and non-perturbative effects\footnote{In (\ref{Psi_UV}) $g$ should be understood
as the running coupling constant $g(\Lambda)$ at some very high scale $\Lambda >> \Lambda_{\rm QCD}$. 
The non-perturbative mass
parameter $m$ should be related to $g(\Lambda)$ and $\Lambda$ via $m\sim \Lambda e^{-8\pi^2/{b_0 g^2}}$.}, such
terms will not spoil the standard results of the QCD perturbation theory.
In essence, what we propose is to {\it assume} that such terms can be summed up to
give a very particular form for the kernel, motivated by our results regarding
the $2+1$ YM theory as well as the existing lattice simulations concerning the
spectrum of $3+1$ YM theory.

We now proceed with a very particular kernel discussed in the section 2.2 for the
case of $b=2/3$. To motivate this choice for the parameter $b$ we may naively argue as follows:
In $2+1$ dimensions the magnetic field $B$ that features in the vacuum wave functional
is a scalar. In $3+1$ dimensions the magnetic field $B_i$ is a vector, so we naturally have
to contract this vector index in
the vacuum wave functional above.
By averaging the scalar product of 3 vectors $v^i$ in 3d over all directions 
(a fixed direction being defined by the usual Euler angles $\theta$ and $\phi$) one gets
$\frac{4\pi}{3} v^2$,
where $\frac{4\pi}{3}$ comes from the angular integral $\int {\cos^2{\theta}} \sin{\theta} d\theta d \phi$.
Per $2\pi$ range of the $\phi$ angle we also get the factor $2/3$. 
Granting its very naive nature, this might be a 
heuristic way to understand why the $2/3$ factor seems to be ``natural'' in the
discussion of the vacuum wave functional for $3+1$ YM given
its $2+1$ dimensional counterpart, but we admit that this 
simplistic argument is too loose and therefore only a phenomenological justification
is possible at present.

Thus we take the relevant kernel from section 2.2
by choosing $b=2/3$
\begin{equation}
K^{-1}(L) =  {\sqrt{L}}\frac{J_{2}(6 \sqrt{L})}{J_{3}(6 \sqrt{L})}.
\end{equation}
The constituent masses are controlled by the zeros of $J_3$,
\be
M_n = \frac{ \gamma_{3,n} m}{3}
\ee
Note that this is different from the usual $2+1$ result
$M_n = \frac{ \gamma_{2,n} m}{2}$, which involves zeros of $J_2$!
In order to determine the mass gap (or the mass of the lowest lying  
glueball) consider the $0^{++}$ states which may be probed
by the operator $Tr B_i^a B_i^a$. We have
\begin{equation}
\langle {{\rm Tr}\,(B_i^a B_i^a)}_x \
{{\rm Tr}\,(B_i^a B_i^a)}_y\rangle \sim \left( K^{-1}(|x-y|)\right)^2.
\end{equation}
The correlation function follows by Wick theorem
given our knowledge of the quasi-gaussian vacuum wave functional, in complete  
parallel with \cite{prl}.

Now, following the general procedure outlined in
section 2 the mass gap, or the mass of $0^{++}$ is given as a sum of two constituents
\begin{equation}\label{eq:0++}
M_{0^{++}} = M_1 + M_1
\end{equation}
and 
\begin{equation}\label{eq:0++*}
M_{0^{++*}} = M_1 + M_2
\end{equation}
etc, as well as
\begin{equation}\label{eq:2++}
M_{2^{++}} = M_2 + M_2
\end{equation}
etc, as in \cite{prl}. Other states follow as in \cite{prl}.
Note that that relevant zeros of $J_3$  are \cite{bessel}:
\begin{equation}
\gamma_{3,1} = 6.380, \gamma_{3,2} = 9.761, \gamma_{3,3} = 13.015, 
\gamma_{3,4} = 16.223, \gamma_{3,5} = 19.409
\end{equation}
Given the phenomenological
nature of our discussion, we note that the zeros of the $J_3$ Bessel function 
do seem to provide the best fit to the available lattice data.

In order to be able to compare our results for glueball masses with 
actual lattice data, we have to re-express them in terms of some other
measurable quantity\footnote{However, we would like to point out that eqs.(\ref{eq:0++})-(\ref{eq:2++}) can readily be used
to obtain definite numerical values for ratios of glueball masses.} like $\sqrt{\sigma}$ or $\Lambda_{\rm QCD}$.
In other words, we need to establish the relationship between the string tension and
the mass parameter $m$ similar to eq.(\ref{eq:stringtension}) in (2+1)D. At the moment we do not
have any definite proposal on the possible generalization of that equation
to (3+1)D. To circumvent this difficulty we should consider the coefficient in (\ref{eq:stringtension}) as
an extra fit parameter. Surprisingly enough, the best fit value for this 
parameter turns out to be very close to $\sqrt{\pi\over 2}$. For this reason in what follows we
simply use the 2+1 dimensional result (\ref{eq:stringtension}) to express glueball masses in terms of the
square root of the string tension.    


We summarize the actual numerical results in a couple of tables.
As noted in the introduction, our analysis can be done
at any rank N, but we do comparisons with the large N data only.
The large N data are from \cite{Lucini:2004my,T}.
There are also many more data available for $SU(3)$ and $SU(8)$.
The large N extrapolation is done by multiplying these
data by $\sqrt{N^2-1}/N$, as we did in $2+1$.
The theoretical results are shown as well as the respective 
difference from the lattice data.
As one can plainly see that the zeros of $J_3$ work very well.

\begin{table}
\caption{\label{Table1} Large N glueball masses in $YM_{3+1}$. All masses are in units of the square root of the string tension.
Comments on the origin of the lattice data are presented in Section 4.1}
\begin{tabular}{l|cccc}
State & Lattice, $N\to\infty$ & Theory  & Difference in \%\\
\hline
$0^{++}$ & $3.307 \pm 0.053$ & $3.394$  & $[2.6]$\\
$0^{++*}$ & $4.67 \pm 0.29$ & $4.294$  & $[8]$\\
$2^{++}$ & $4.80 \pm 0.14$ & $5.193$  & $[8]$\\
$2^{++*}$ & $6.17 \pm 0.21$ & $6.058$ & $[2]$\\
$4^{++}$ & $7.81 \pm 0.20$& $6.924$  & $[11]$\\
$4^{++*}$ & $9.88 \pm 0.36$& $7.779$  & $[21]$\\
$6^{++}$ & $9.34 \pm 0.55$ & $8.634$  & $[7.5]$\\
\end{tabular}
\end{table}

\subsection{Comments: lattice vs. theory}

The only genuine large-$N$ lattice estimates are available for three states only
\begin{equation}
\frac{m_{0^{++}}}{\sqrt{\sigma}}= 3.307(53)
\end{equation} 
\begin{equation}
\frac{m_{2^{++}}}{\sqrt{\sigma}}= 4.80(14)
\end{equation}
\begin{equation}\label{inverted}
\frac{m_{0^{++*}}}{\sqrt{\sigma}}= 6.07(17)
\end{equation}
This is from the most recent study by Lucini, Teper and Wenger \cite{Lucini:2004my} and is based 
on the extrapolation of results available for N=2,3,4,6,8.
These results are consistent with a previous study by Lucini and Teper \cite{T}(based on N =2,3,4,5). 
The only difference is that error bars are smaller 
in this new simulation (about 1.5 to 3\% as opposed to 5-8\% in Lucini and Teper.)

The $0^{++}$ and $2^{++}$ mass estimates seem to be reliable. They show a smooth approach 
to the continuum limit for any given value of $N$ as well as a smooth 
extrapolation from finite $N$ values to $N=\infty$.

The $0^{++*}$ state is more troublesome.  So we should compare critically data 
from other sources as well. Other sources are the
PhD thesis of H. Meyer \cite{Meyer:2004gx}
(this includes
SU(3) and SU(8) data; somewhat surprisingly, Meyer's data differs from Teper et al.) and Morningstar and
Peardon \cite{Morningstar:1999rf} (SU(3)only; this is considered, at the moment,
to be the most precise lattice study for $N=3$).
First, all these authors report that the finite volume corrections are 
still important for this state and so its identification as a glueball
is questionable (i.e. this state might be a torelon). Second, according to 
Teper {\it et al.} $m_{0^{++*}}\approx 2\ m_{0^{++}}$, which leads to a possibility that 
this state is actually
a two-glueball state with zero relative momentum. Third, the extrapolation of finite N values to $N=\infty$ 
is not so good (as can be seen, for example, from the
inverted slope of the large-$N$ fit in (\ref{inverted})) which suggests the possibility 
that different mass eigenstates were used while performing the fit to $N=\infty$.
Finally, there is a large discrepancy between Teper's $N=\infty$ result  
$m_{0^{++*}}\approx 2\ m_{0^{++}}$ and the SU(3) result by Morningstar and Peardon
$m_{0^{++*}} = 1.54\ m_{0^{++}}$  or the $SU(8)$ result by H.~Meyer $m_{0^{++*}} = 1.42\ m_{0^{++}}$.
   
For these reasons we think that $0^{++*}$ mass estimate by Teper {\it et al.} might be doubted. A more 
reasonable estimate of the $0^{++*}$ mass can be based
on the $SU(8)$ value 
found in Meyer's thesis, which upon rescaling leads to our $N=\infty$ estimate of $0^{++*}$ mass 
\begin{equation}
\frac{m_{0^{++*}}}{\sqrt{\sigma}}= 4.67 \pm 0.29
\end{equation}
This is the value we use in Table 1 and it's compatible with Morningstar and 
Peardon (i.e. approximately 1.5 times heavier than $0^{++}$ mass).
However, we would like to point once again that this estimate should be 
treated with caution. Similarly, the rest of the lattice values in Table 1
(for $2^{++*}$, $4^{++}$, $4^{++*}$, $6^{++}$ states) were obtained from $SU(3)$ and $SU(8)$ 
data in \cite{Meyer:2004gx} by simple rescaling and therefore should
also be treated with caution.

Note that Meyer's thesis contains $SU(3)$ estimates for the
masses of $0^{++**}$ and $0^{++***}$. However, 
these seem to be too heavy to be absolutely trusted (they are in the same ballpark as
Teper's $0^{++*}$ estimate) and therefore we 
choose not to include them in our comparison.

This concludes our comparison in the $PC=++$ sector of the theory and we want 
to proceed now to the $PC=-+$ sector. But before we do
that we would like to make a few general comments on the qualitative properties of glueball 
mass spectrum in both $2+1$ and $3+1$ dimensions.
It has been known for a while \cite{constituent,Kuti:1998rh} that the low mass glueball 
spectrum can be qualitatively understood in terms of local gluon interpolating
operators of minimal dimension: higher dimensional operators create higher mass states. In 
particular, in $2+1$ dimensions there is a (mass) dimension-$6$ operator $tr(B^3)$ with $0^{--}$ quantum
numbers and this is consistent with the lattice result concerning the
existence of a light (with $m_{0^{--}}\approx 1.5 m_{0^{++}}$) $0^{--}$ glueball state in the 
spectrum of pure Yang-Mills theory in $2+1$
dimensions. On the contrary, in $3+1$ dimensions modern lattice studies seem to strongly 
rule out the existence of light $0^{--}$ glueball states \footnote{The only definite
estimate of $0^{--}$ mass, $m_{0^{--}} = (2.44 \pm 0.25) m_{0^{++}}$ , can be found 
in \cite{Hu:1996ys} and is indeed quite heavy.}. Qualitatively this can be explained by the fact
that dimension-$6$ interpolating operator with $0^{--}$ quantum numbers simply does not exist in $3+1$D. 

What does exist in $3+1$D however is ${\cal O}_E = f_{abs} (\vec E^a \times \vec E^b) \cdot \vec E^c$ 
operator with $0^{-+}$ quantum numbers
\footnote{In fact, in $3+1$D it is possible to construct also a
dimension-$4$ interpolating operator with $0^{-+}$ quantum numbers, which is 
$\vec E^a \cdot \vec B^a$. Similarly, one can construct $0^{++}$ operators $(\vec E^a)^2 -(\vec B^a)^2$ 
and
$f_{abc} (\vec B^a \times \vec B^b) \cdot \vec B^c$ of dimension four and six respectively. However 
as was argued previously in \cite{constituent}, for a broad
class of constituent glue models the lightest $0^{-+}$ glueball couples to dimension-$6$ operator 
only. This explains our choice of 
${\cal O}_E = f_{abs} (\vec E^a \times \vec E^b) \cdot \vec E^c$ as a probe
operator for this state.}. It should be clear now that in complete analogy with \cite{prl}, i.e. by 
considering an equal-time correlator of two such
operators $\langle {\cal O}_E(x)\,{\cal O}_E(y) \rangle$, we should think of 
glueball mass spectrum in $PC=-+$ sector
as being constructed out of three constituent masses. More specifically, we obtain
\begin{eqnarray}
M_{0^{-+}} = M_1 + M_1 + M_1 = 5.091\, \sqrt\sigma\\
M_{0^{-+*}} = M_1 + M_1 + M_2 = 5.99\, \sqrt\sigma\\
M_{2^{-+}} = M_1 + M_2 + M_2 = 6.89\, \sqrt\sigma\\
M_{2^{-+*}} = M_1 + M_2 + M_3 = 7.76\, \sqrt\sigma
\end{eqnarray}

A comparison of these predictions with lattice data is presented in Table 2. No 
genuine large $N$ lattice data are available for $PC=-+$ states. Therefore,
as before, our large $N$ "lattice" estimates were obtained by 
a simple rescaling of the actual $SU(3)$ and $SU(8)$ lattice data presented in \cite{Meyer:2004gx}.  

Finally, we include two comments indicating
potential difficulties in our scheme: 

1. As can be seen from Table 2 the seeming large discrepancy 
for the masses of $0^{-+*}$ and $2^{-+}$ glueballs can be avoided if we swap theoretical
values for these states. After all, there is no particular reason why $M_1 + M_1 + M_2$ combination 
should be associated with $0^{-+*}$ and not with
$2^{-+}$. Same for $M_1 + M_2 + M_2$ combination. 
On the other side this is similar to the situation 
involving $0^{++*}$ and $2^{++}$ glueballs: here the lattice data favor $m_{0{++*}} > m_{2^{++}}$.
(However, the final status of the $0^{++*}$ mass on the lattice is far from clear.)
Nevertheless, our model predicts otherwise.

2. The $0^{-+}$ mass agrees nicely with the lattice
result. However, there is a claim in the lattice literature that $0^{-+}$ should be
heavier than $2^{++}$. This can not be seen from large-N extrapolated data in Tables 1 
and 2 (they seem to be simply degenerate) but Morningstar
and Peardon say that ``the pseudoscalar is clearly resolved at $7\sigma$ level to be heavier than 
the tensor''. Our current theoretical model (based on 
zeros of $J_3$) predicts just the opposite (compare Tables 1 and 2). It's easy 
to see however that $M_{0^{-+}} > M_{2^{++}}$ requirement translates into
$\gamma_{\nu,2} < {3\over 2}\ \gamma_{\nu,1}$ requirement for Bessel zeros. This 
is possible if $\nu > 3.3$. Therefore, from this point of view
$\nu = 3$ is not completely satisfactory!

\begin{table}
\caption{\label{Table2} Large N glueball masses in $YM_{3+1}$. All masses are in units of the square root of the string tension.
Comments on the origin of the lattice data are presented in Section 4.1.}
\begin{tabular}{l|cccc}
State & Lattice, $N\to\infty$ & Theory  & Difference in \%\\
\hline
$0^{-+}$ & $4.81 \pm 0.13$ & $5.091$  & $[6]$\\
$0^{-+*}$ & $7.22 \pm 0.33$ & $5.99$  & $[22]$\\
$2^{-+}$ & $5.95 \pm 0.10$ & $6.89$  & $[15]$\\
$2^{-+*}$ & $7.46 \pm 0.29$ & $7.76$ & $[4]$\\
\end{tabular}
\end{table}

In spite of these important caveats we take the above phenomenological
fit to the lattice data 
as a positive indication that the general form of the wave functionals we have discussed in this note
does capture the essential non-perturbative physics.
Obviously, we are only scratching the surface in our understanding of what is truly
going on here. Much more detailed work is needed to elucidate the real nature of the
wave functionals discussed in this paper.

\vskip 1cm

{\bf \Large Acknowledgements}

\vskip .5cm

We would
like to thank  B. Bergoltz, D. Karabali, V. P. Nair, P. Orland, R. Pisarski,
S. Rajeev and other participants of the City College workshop on non-perturbative Yang Mills theory,
for
interesting conversations and questions regarding the material presented in this  
paper. We also thank various participants of the Virginia Tech first Sowers
workshop in theoretical physics for questions and comments.
{\small RGL} was supported
in part by the U.S. Department of Energy under contract
DE-FG02-91ER40709. {\small DM}
is supported in part by the U.S. Department of Energy
under contract DE-FG05-92ER40677. Many thanks to Perimeter Institute, the Kavli Institute
for Theoretical Physics, and IHES at Bures sur Yvette, for their respective hospitalities.

\end{document}